\documentclass{preprint}
\usepackage{epsfig,graphics}
\volume{0}     
\pubyear{2000}
\firstpage{0} 
\lastpage{4}  
\begin{document}

{\ }
\vspace{10cm}
\begin{center}
This is an unedited preprint. The original publication is available at \\ {\ } \\
http://www.springerlink.com \\ {\ } \\
http://www.doi.org/10.1023/A:1012606309138
\end{center}
\newpage

\begin{frontmatter}  
\title{Interpretation of M\"ossbauer spectra in the energy and time domain with neural networks}       
\author[A]{H. Paulsen}      
\author[B]{R. Linder}
\author[C]{F. Wagner}
\author[A]{H. Winkler}
\author[B]{S.J. P\"oppl}
\author[A]{A.X. Trautwein}
\address[A]{Institut f\"ur Physik, Med. Universit\"at L\"ubeck, D-23538 L\"ubeck, Germany 
                  \email{paulsen@physik.mu-luebeck.de} }
\address[B]{Institut f\"ur Med. Informatik, Med. Universit\"at L\"ubeck, D-23538 L\"ubeck, Germany}
\address[C]{Institut f\"ur Theoretische Physik, Christian-Albrechts-Universit\"at, D-24118 Kiel, Germany}
\runningauthor{H. Paulsen et al.}
\runningtitle{Interpretation of M\"ossbauer spectra with neural networks}
\begin{abstract}  
An artificial neural network for extracting reasonable and fast estimates of hyperfine parameters from M\"ossbauer spectra in the energy or time domain is outlined.
First promising results for determining the asymmetry of the electric field  gradient at the nucleus of a diamagnetic iron
center as derived with different types of neural networks are reported
\end{abstract}
\begin{keywords}  
M\"ossbauer spectroscopy, nuclear forward scattering, neural network
\end{keywords}
\classification{} 
\end{frontmatter}

\renewcommand{\baselinestretch}{1.3} \Large \normalsize

Artificial Neural networks (ANN) are used for the analysis of experimental data in a broad range of scientific disciplines.
In M\"ossbauer spectroscopy ANN have been used to detect corrosion products \cite{R1} or to establish databases
of known spectral patterns \cite{R1a}.
In the work outlined here ANN are used to build a program package that allows to obtain a reasonable and fast estimate of
hyperfine parameters, e.g. the isomer shift, the electric field gradient tensor, the magnetic hyperfine interaction
tensor, from experimental M\"ossbauer transmission spectra (MTS) or nuclear forward scattering (NFS) spectra.
The fast estimate of these parameters is obligatory when a fast decision is required whether and under which conditions
(temperature, applied field) further measurements have to be performed, especially if time is a limiting factor like for recording
NFS spectra at 3$^{\rm rd}$ generation machines.
Additionally, such an estimate is useful in providing a set of start parameters for, e.g., a conventional spin-Hamiltonian analysis
of the experimental spectra \cite{R5,R6}.
In a first step, that is reported here, only the determination of the asymmetry parameter $\eta$ of the electric field  gradient at the iron nucleus of a diamagnetic and randomly oriented complex is considered.

The development of this program package includes an ANN with supervised learning and consists of three parts:
(i) choosing appropriate network architectures and learning rules, (ii) training the ANN, and (iii) testing the performance of the ANN with a set of spectra that have not been used for training.
In the present study we have used feed-forward networks \cite{R7} with one or two hidden layers\footnote{One hidden layer is necessary and sufficient to overcome the limitations of a perceptron, however two hidden layers might be more efficient in some cases.}
of neurons and two different learning rules, i.e. an improved backpropagation algorithm (APROP) \cite{R3} and, included in the 
NETFIT program package
\cite{R4a}, a variable metric minimization algorithm (MINIM) \cite{R4}.
The ANN was realized as C++ (APROP) or Fortran (NETFIT) programs, respectively. Each neuron is realized by a set of variables containing (i) the addresses of the connected neurons, (ii) for each connection a weight, that is determined during the training process, and (iii) the current value of the neuron representing a data point of the measured spectrum (input layer), the retrieved
parameter (output layer), or data of no obvious physical meaning (hidden layers).
To train the ANN 5000 M\"ossbauer transmission spectra have been simulated using the spin-Hamiltonian formalism \cite{R5}
together with powder averaging.
While the linewidth $\Gamma=0.37$ mm/s was kept fixed for all spectra, the quadrupole splitting $\Delta E_Q$, the asymmetry parameter $\eta$, and the external field $B$ were chosen at random from the intervals [1$\ldots$4 mm/s], [0$\ldots$1], and [4$\ldots$8 T], respectively.
Noise corresponding to an off-resonance count rate of 10$^6$ was introduced using random numbers with Poisson distribution.
As input for the ANN the normalized transmission count rates in 256 channels covering the velocity range from -8 to +8 mm/s were used.
In the same way NFS spectra have been simulated using the SYNFOS program \cite{R6} with a simulated noise level that
corresponds to a maximum count rate of 1000.

During training each ANN output, $\eta_{\rm net}$, was compared with the actual value $\eta$ of its corresponding input spectra.
The sum $\chi^2(T)$ of the squared differences,
\begin{equation}
  \chi^2(T) = \frac{1}{|T|} \sum_{s \in T} |\eta(s)-\eta_{\rm net}(s)|^2 \quad ,
\end{equation}
where the index $s$ enumerates the 5000 different spectra of the training set $T$, is a measure for the memorization ability of the ANN.
$\chi^2(T)$ was minimized iteratively by changing as internal ANN parameters the weights $w_{ij}$ between neurons $i$ and $j$ of adjacent layers.
In case of APROP supervision of the training procedure demands a second, independent set of 1000 spectra, which was generated in
the same way as the training set.
After training was completed the ANN was kept fixed and its performance was tested by an additional independent set G containing
1000 spectra.
The performance, or more precisely, the generalization ability of the trained ANN is represented by $\chi^2(G)$ or, alternatively, by the
probability to obtain an ANN output $\eta_{\rm net}$ with less than 0.05 absolute deviation from the actual value $\eta$,
\begin{equation}
  p_{0.05} = \frac{1}{|G|} \sum_{s \in G} \Theta(0.05-|\eta(s)-\eta_{\rm net}(s)|) \quad .
\end{equation}

\begin{table}[t]
\centering
\caption{Performance of different ANNs for M\"ossbauer spectra in energy (MTS) and time domain (NFS).}\label{tab1}
\tabcolsep=5pt
\begin{tabular}{llcccccccc}
\hline
 Spectra & Preprocessing &      Hidden       & learning && \multicolumn{2}{c}{Training} && \multicolumn{2}{c}{Generalization} \\[0.2ex]
\cline{6-7}\cline{9-10}
\multicolumn{10}{c}{}\\[-2.5ex]
              &                            &      neurons     & rules       &&   $\chi^2$   &   $p_{0.05}$   &&      $\chi^2$      &     $p_{0.05}$    \\
\hline
 MTS      & no                        &            3         & MINIM    &&       0.07      &          0.55     &&          0.09        &         0.44          \\
 MTS      & no                        &  2 $\times$ 20 & APROP   &&       0.05      &          0.81     &&          0.08        &         0.57          \\
 MTS      & yes (Lorentzian fit) &          5           & MINIM    &&       0.05      &          0.75     &&          0.06        &         0.71          \\
 MTS      & yes (Lorentzian fit) &  2 $\times$ 20 & APROP   &&       0.04      &          0.82     &&          0.06        &         0.78          \\
 NFS      & no                        &           4          &  MINIM    &&       0.06      &          0.63     &&          0.09        &         0.55          \\
 NFS      & no                        &  2 $\times$ 20  & APROP   &&       0.08      &          0.76     &&          0.06        &         0.65          \\
\hline
\end{tabular}
\end{table}

Both ANN that have been used here independently, NETFIT and APROP, predict $\eta$ from MTS with an
accuracy that should be sufficient in most cases (Tab. 1).
The same accuracy can be reached for the NFS spectra.
In view of the expansion of the program for simultaneously obtaining several hyperfine parameters (and not only one as in the
present case) it is desirable to keep the ANN and the computational effort during the training as small as possible.
For this reason preprocessing of the M\"ossbauer spectra was introduced, i.e. before starting the ANN procedure the spectra were fitted by six Lorentzians with fixed linewidth.
The position and the height of these Lorentzians together with $\Delta E_Q$ and the external field $B$ are then presented to the ANN.
As a result the performance, as measured by $\chi^2$ and $p_{0.05}$, improves (Tab. 1, Fig. 1).
At the same time the number of neurons in the input layer is reduced from 256 to 14.
In summary we state that a relatively small ANN with 14 input neurons, 5 hidden neurons, and 1 output neuron provides a reasonable estimate for $\eta$.
This situation is promising in view of expanding the ANN procedure towards a fast and simultaneous multi-parameter analysis of M\"ossbauer spectra.

\begin{figure}
\hspace{\fill}
\scalebox{0.38}{\includegraphics{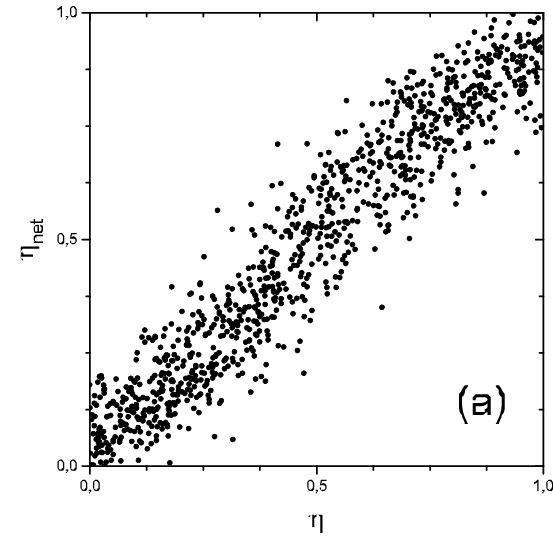}} \hspace{\fill}
\scalebox{0.38}{\includegraphics{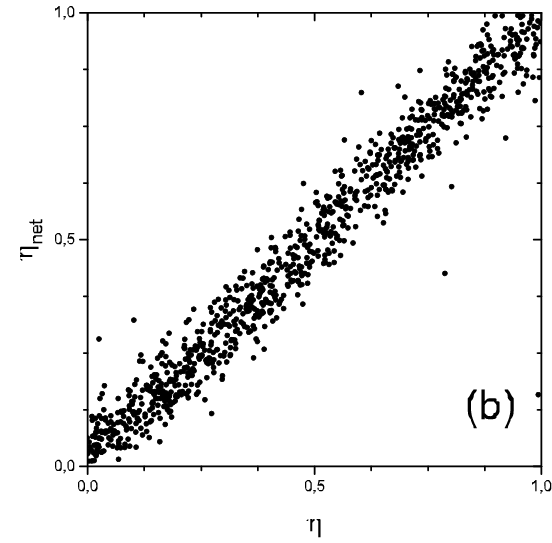}}
\hspace{\fill}
\caption{\label{fig1} Generalization ability: ANN output $\eta_{\rm net}$ vs the corresponding actual $\eta$ for 1000 MTS spectra in set $G$. In (a) raw data and in (b) Lorentzian fits were used as ANN input. } 
\end{figure}

\end{document}